\documentclass[aps,12pt,tightenlines,amsmath,amssymb,showpacs]{revtex4}
\usepackage{graphicx}
\usepackage{bm}
\newcommand{\be}{\begin{equation}}\newcommand{\ee}{\end{equation}}
\def\obsrep{observable representation} \def\Obsrep{Observable representation}
\def\Acal{{\cal A}}
\def\erset{\mathbb R}
\def\Abf{{\bf A}}
\def\parsup#1{^{(#1)}} % PARenthetical SUPerscript
\def\Rcg{\widetilde R}
\def\KK{Kobe and Krawczyk}
\def\prlsection#1{\bigskip\noindent\textit{#1}.--- }
\def\x{$\mskip -1.5mu\times\mskip -1.5mu$}

\begin{document}
%%%%%%%%%%%%%%%%%%%%%%%%%%%%%%%%%%%%%%%%%%%%%%%%%%%%%%%%%%%%%%%%%%%%%%
\title{Mean Field Spin Glass in the Observable Representation}
\author{L. S. Schulman}
\affiliation{Physics Department, Clarkson University, Potsdam, New York 13699-5820, USA}
\email{schulman@clarkson.edu}
\date{\today}
%%%%%%%%%%%%%%%%%%%%%%%%%%%%%%%%%%%%%%%%%%%%%%%%%%%%%%%%%%%%%%%%%%%%%%
\begin{abstract} 
The state space for the $N$-spin mean field (SK) spin glass---nominally an $N$-cube---is embedded in a low dimensional continuous space in such a way that metastable and stable phases can easily be discerned, a concept of nearness of configurations defined, and peaks in the Parisi $q$-parameter overlap distribution identified. The dynamical and partly hierarchical interrelation of these phases can be directly imaged.
\end{abstract}
%%%%%%%%%%%%%%%%%%%%%%%%%%%%%%%%%%%%%%%%%%%%%%%%%%%%%%%%%%%%%%%%%%%%%%
\maketitle
%%%%%%%%%%%%%%%%%%%%%%%%%%%%%%%%%%%%%%%%%%%%%%%%%%%%%%%%%%%%%%%%%%%%%%

The mean field (SK) spin glass \cite{sherrington} continues to be an active focus of research both for its own sake and for the light it can cast on the short-range spin glass \cite{guerra, janke, billoire, berg, aizenman, marinari}. Nevertheless, problems remain. Among these is the difficulty of visualizing its state space. One would like to have an idea of the energy landscape, the absolute and local minima, an image of what states are ``near'' those minima and perhaps even get a handle on the elusive notion of pure state in finite systems. It would be especially convenient if phases, metastable or stable, could be identified and well-localized on this landscape. A geometric picture in which at least some of these goals can be achieved is the object of the present article.

For systems evolving under stochastic dynamics there is an embedding, known as the \textit{observable representation} \cite{multiplephases, imaging}, of the state space into continuous spaces of various dimension. In it the features just mentioned stand out in simple form, with phases identified as the extrema of a certain convex set and with a natural distance inherited from the dynamics on the system. This provides a direct image of the interrelationship between phases and of the hierarchical structure of passage from phase to phase. The notion of metastable phase---problematic in infinite volume systems---is completely natural in this context. Finally, there is no restriction of our method to the SK model, the only issue being whether lattice models exhibit structure for systems small enough for our technique to be applied.

The \obsrep\ provides an abstract definition of phase (stable and metastable) and can be used to see the relation of phases to the Parisi overlap parameter \cite{parisi}, ``$q$.''  In addition the temporal flow of child phases into parents can be displayed. In this article I use as the state space all $2^N$ configurations and relatively simple matrix methods. As I will also show, considerable reduction in the state space is possible, opening the door to extensive use of this method.

I first recall the definition of the \obsrep. Then I offer evidence for structure in the 12-spin system (those less troubled by state-space size have not looked at such small systems). Then I get to the main point, which is to show how structure is displayed in the \obsrep. I close with a discussion of prospects.

%%%%%%%%%%%%%%%%%%%%%%%%%%%%%%%%%%%%%%%%%%%%%%%%%%%%%%%%%%%%%%%%%%%%%%
\prlsection{The observable representation and associated notation}
%%%%%%%%%%%%%%%%%%%%%%%%%%%%%%%%%%%%%%%%%%%%%%%%%%%%%%%%%%%%%%%%%%%%%%
States are $x,y,z\in X$, with $X$ a finite set. The system moves in discrete time according to an (assumed) irreducible stochastic matrix $R$: $ R_{xy}= \Pr\bigl(z(t+1)=x\bigm| z(t)=y\bigr)$. For a spin glass, $R$ satisfies detailed balance (although Refs.\ \cite{multiplephases,firstorder} are more general). The eigenvalues of $R$ are therefore real and can be written $1=\lambda_0 >|\lambda_1| \geq|\lambda_2|\dots\geq0$. $\lambda_0$ is associated with the stationary distribution: $Rp_0=p_0$; the eigenvalue relation for the associated left eigenvector, $A_0(x)\equiv1$, expresses conservation of probability. Left eigenvectors are designated $A_k$ (same $k$ as in $\{\lambda_k\}$) and right eigenvectors $p_k$. They satisfy $\langle A_k|  p_j\rangle = \delta_{kj}$ and are normalized by $\max_x |A_k(x)|=1$.

The $m$-dimensional \textit{observable representation} is the following set (to be thought of as an embedding of $X$ in $\erset^m$)
\be
\Acal_m\equiv \left\{ {\Abf}\in\erset^m
\mid \Abf=(A_1(x),\dots,A_m(x)) \,,\; x\in X\right\}
\label{eq:observrep} \,.
\ee
To visualize, write the first $m$ left eigenvectors as row vectors, one atop the other. The points of $\Acal_m$ are then the \textit{columns}:
\be
\begin{array}{cccccc}
        &    &\Abf(x_1)  &\Abf(x_2)    &\Abf(x_3)  &\dots  \cr
        &    &\downarrow &\downarrow   &\downarrow &       \cr        
 A_1\to &~~  &A_1(x_1)   &A_1(x_2)     &A_1(x_3)   &\dots  \cr
 A_2\to &    &A_2(x_1)   &A_2(x_2)     &A_2(x_3)   &\dots  \cr
\vdots  &    &\vdots     &\vdots        &\vdots    &       \cr 
A_m\to  &    &A_m(x_1)   &A_m(x_2)     &A_m(x_3)   &\dots  
\end{array}
\label{eq:visualization}
\ee
In \cite{multiplephases} we established the properties of $\Acal_m$ when there is a phase transition, which  corresponds \cite{firstorder} to eigenvalue near-degeneracy for the largest eigenvalues. If $\lambda_m$ is close to 1, while $\lambda_{m+1}$ is not, $\Acal_m$ is a simplex with points belonging to the phases located at the vertices. Interior points are not in phases, but their barycentric coordinates with respect to the extrema are the probabilities that these points evolve towards the corresponding phase.

In \cite{multiplephases} we distinguished which assertions remain true if $\lambda_{m+1}$ is not small. The convex hull of $\Acal_m$ need not be a simplex, but there still are extrema, and points dynamically close and in a phase cluster about the extrema. The relation between dynamical proximity and closeness in $\Acal$ also holds for non-extrema \cite{imaging}. Let $D(x,y;t)\equiv \sum_u\left|(R^t)_{ux}-(R^t)_{uy}\right|/\sqrt{p_0(u)}$. $(R^t)_{ux}$ is the probability that starting at $x$ one arrives at $u$ in $t$ time steps. Then $D(x,y;t)/|\lambda_m|^{t} \geq \left[\sum_{\alpha=1}^m \left|A_\alpha(x) -A_\alpha(y)\right|^2\right]^{1/2}$. Points whose distributions merge are spatially close in~$\Acal$. 

%%%%%%%%%%%%%%%%%%%%%%%%%%%%%%%%%%%%%%%%%%%%%%%%%%%%%%%%%%%%%%%%%%%%%%
\prlsection{ Small mean-field models}
%%%%%%%%%%%%%%%%%%%%%%%%%%%%%%%%%%%%%%%%%%%%%%%%%%%%%%%%%%%%%%%%%%%%%%
I next show that for $N$ as small as 12 there are structures associated with slow relaxation to more stable phases. The energy is $E=-\sum_{j,k=0}^NJ_{jk}\sigma_j\sigma_k/2\sqrt{N}$, with $J_{jk}=\pm1$ the quenched bonds and $\sigma_{j}=\pm1$ the spins. The temperature-$T$ transition probability from $x\equiv (\sigma_1,\sigma_2, \dots,\sigma_N)$ to $x'\neq x$ is $R(x',x)= \exp\left[-\left( E(x')-E(x) \right)/T\right]/N$ when $E(x')>E(x)$; otherwise it is $1/N$. I permit only single spin flips.

Structure was observed in the time-dependent distribution of the Parisi overlap parameter, ``$q$'' \cite{parisi, berg, billoire}: For fixed $J$, two random initial conditions ($x\parsup{\ell}$) are taken. They evolve separately under the stochastic dynamics and the overlap $q\equiv x\parsup1\!\cdot\! x\parsup2/N \equiv(1/N)\sum_{k=1}^N \sigma_k\parsup1\sigma_k\parsup2$ is calculated. Values of $q$ are collected for many times and many initial conditions. In the distribution function for $q$, peaks represent persistence in pairs of metastable phases, and the time dependence of the distribution reflects relaxation.

\begin{figure}
\centerline{
\includegraphics[height=.2\textheight,width=.3\textwidth]{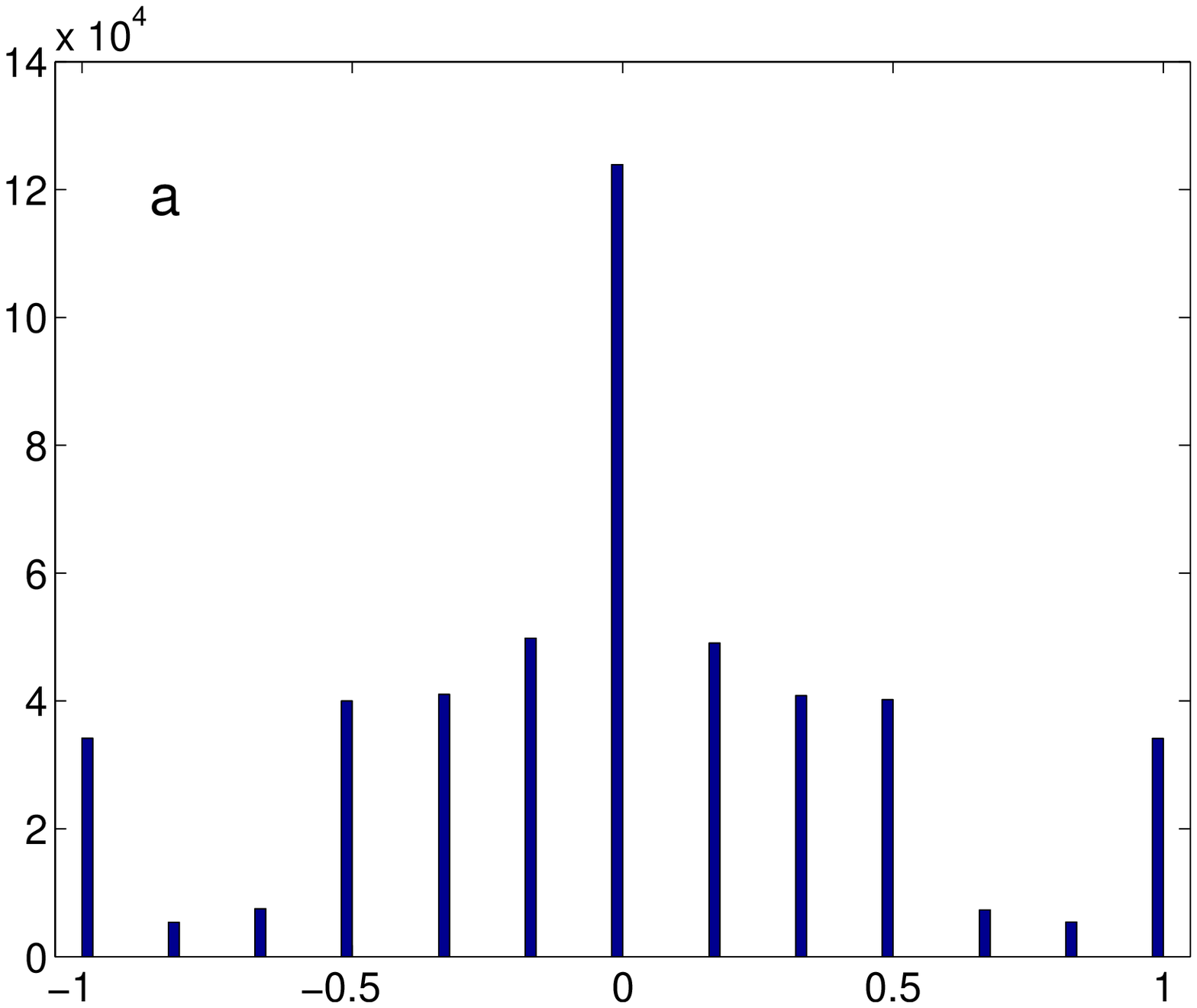}~
\includegraphics[height=.2\textheight,width=.3\textwidth]{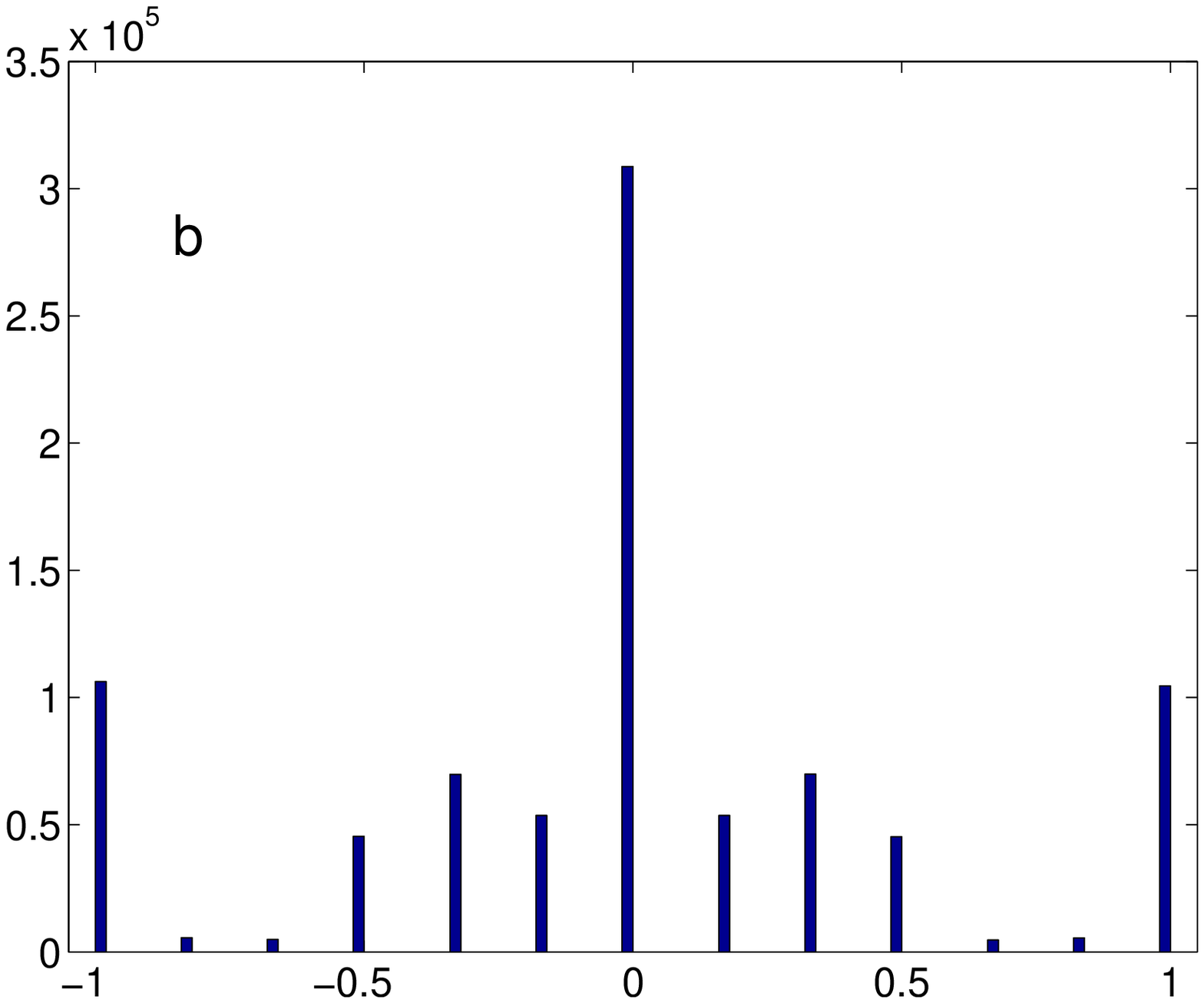}~
\includegraphics[height=.2\textheight,width=.3\textwidth]{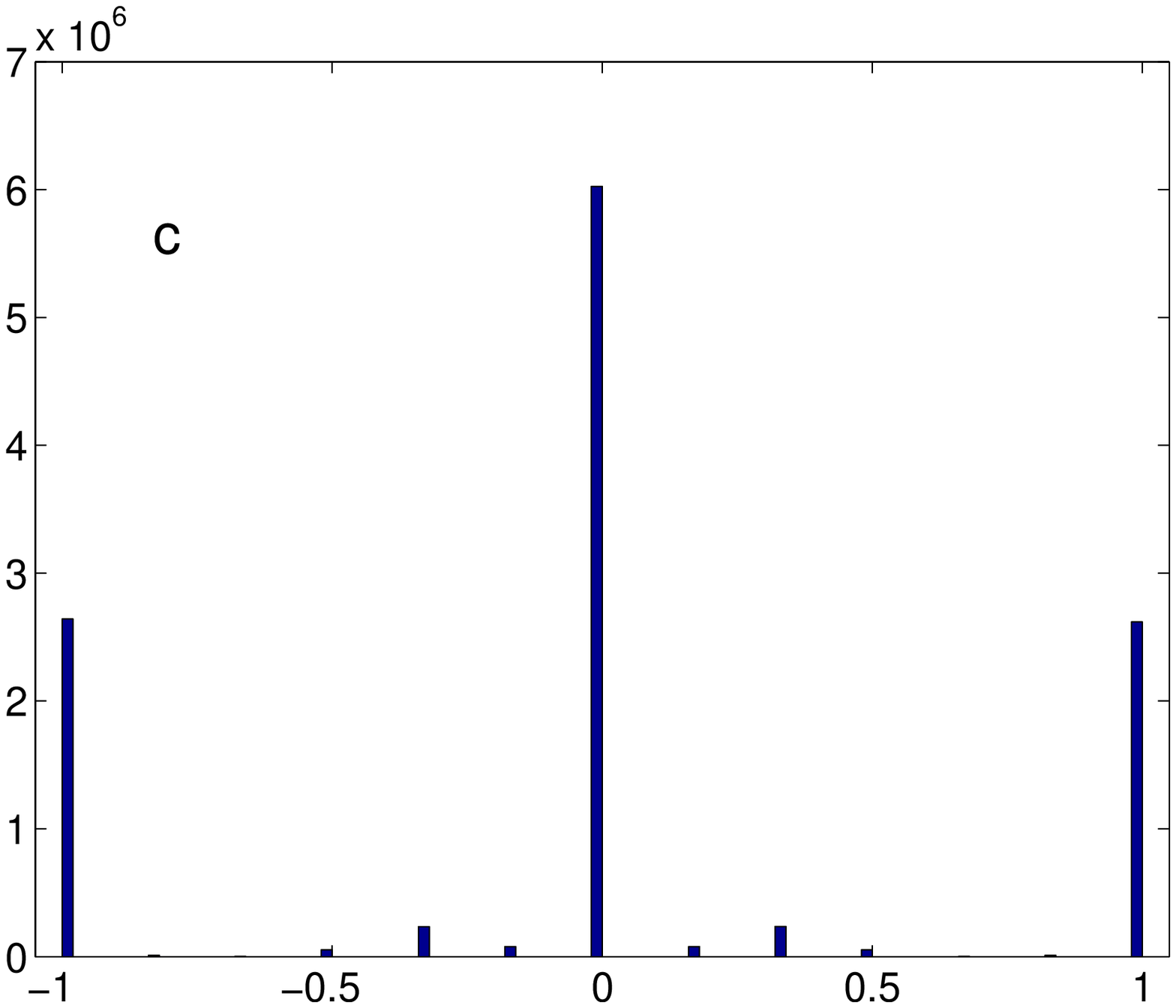}}
\caption{Histogram of the distribution function for ``$q$,'' for the same quench, for a succession of times. $T=0.2\,T_c$ and $t=100$, 200, and 3000. The central peak does not correspond to an absence of ordering. Rather the phases break into pairs, with half the pairwise products, $x\parsup j\cdot x\parsup k$, being zero. Fig.\ \ref{fig:structureq}c is close to the infinite-time distribution, $\sum_{x,y}p_0(x)p_0(y) \delta(q-q_{xy})$.\label{fig:structureq}}
\end{figure}

For $N=12$ there is structure. Fig.\ \ref{fig:structureq} shows histograms for a specific quench, $J_{ij}$ (to be called $\hat J$), for three run times. For a few hundred time steps the system gets caught in metastable phases whose importance gradually lessens (space limitations preclude more figures) until about time 3000. This time scale is confirmed by $\{\lambda_k\}$. Note that the central peak does \textit{not} indicate an absence of structure (see the caption).

\begin{figure}
\includegraphics[height=.33\textheight,width=.45\textwidth]{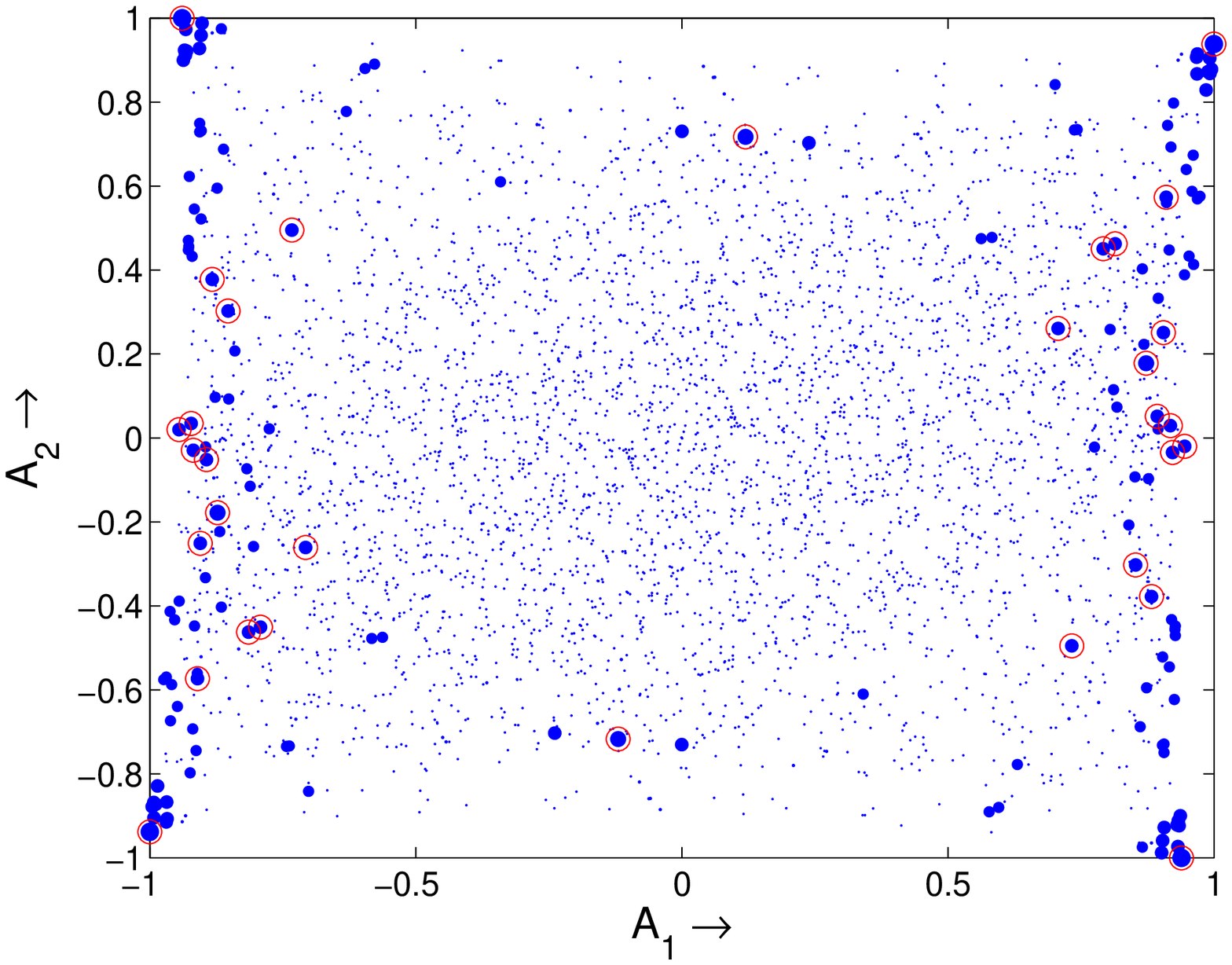 }
\caption{$\Acal_2$ for $\hat J$. Symbol size increases with equilibrium weight (hence lower energy). Circles indicate local minima in energy. \textbf{Note:} \textit{For reasons that are beyond me, the arXiv refuses to print the eps file associated with this figure. Please see the published version of the article.} \label{fig:obsrepJd2}}% \labeld{fig:obsrepJd}}
\end{figure}

\begin{figure}
\includegraphics[height=.33\textheight,width=.45\textwidth]{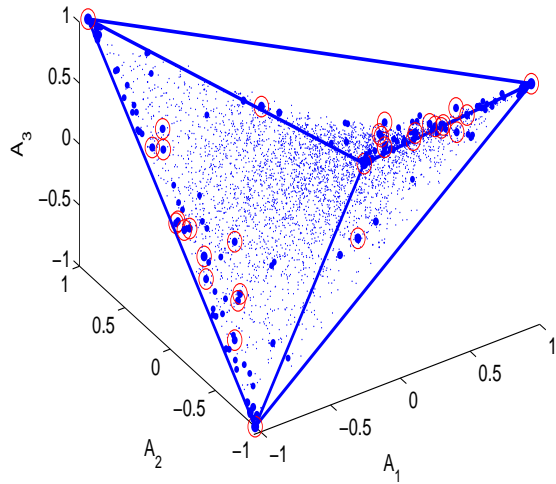}
\caption{$\Acal_3$ for $\hat J$. Symbol size increases with equilibrium weight (hence lower energy). Circles indicate local minima in energy. The convex hull (a tetrahedron) has been outlined. The vertical lines of large-weight points on the left and right sides of $\Acal_2$ are at approximately right angles to one another in $\erset^3$.\label{fig:obsrepJd3}}% \labeld{fig:obsrepJd}}
\end{figure}

%%%%%%%%%%%%%%%%%%%%%%%%%%%%%%%%%%%%%%%%%%%%%%%%%%%%%%%%%%%%%%%%%%%%%%
\prlsection{ Visualizing structure in the \obsrep}
%%%%%%%%%%%%%%%%%%%%%%%%%%%%%%%%%%%%%%%%%%%%%%%%%%%%%%%%%%%%%%%%%%%%%%
Figs.\ \ref{fig:obsrepJd2} and \ref{fig:obsrepJd3} show $\Acal_2$ and $\Acal_3$ for the $R$ associated with $\hat J$. The first few eigenvalues of $1-R$ are: 0, $(2.1,7.1,8.3)\!\times\!10^{-6}$, $(5.58,5.62,11.19,11.20)\!\times\!10^{-4}$, $(7.92,7.93,10.84,10.86)\!\times\!10^{-3}$. By the criteria of \cite{multiplephases}, there are many phases (although for some $J$'s the eigenvalue dropoff is less steep).

The following features can be identified. The four ``corner'' states in $\Acal_2$ are absolute minima. The clusters of nearby points constitute the ground states, and as in \cite{multiplephases} (Fig.\ 3), contain more points than is evident to the eye. The prominent points near $A_1=0$ (at large $|A_2|$) support metastable phases (they are extrema in higher dimension), but do not evolve to unique target phases. A state at one of these positions will ultimately find itself with near 50-50 probability on one side or the other. This statement is not based on simulation (although it can be so verified), but rather on the ``barycentric'' theorem quoted earlier. The prominent points near $A_2=0$ (with large $|A_1|$) have similar properties and asymptotics.

To study the hierarchy of temporal evolution I use the extrema of $\Acal$ to identify phases. For $\hat J$ there are 32 local minima and by $\Acal_7$ they are all extrema. I use them as the nuclei of the phases. Among $J$'s this is a relatively large number, but our results are unchanged when non-local-minimum-extrema are used as nuclei. A significant aspect of phase identification is the positing of a distance on $X$. We have previously discussed several metrics \cite{grains, dynamicalmetric} and I here use $ d_t^2(x,y) \equiv  \sum_{k\geq1} \left|A_k(x)-A_k(y)\right|^2 |\lambda_k|^t $. $t$ is an adjustable time scale. The results are not sensitive to $t$, to the power of $ \left|A_k(x)-A_k(y)\right|$, or to which decreasing (as $k\uparrow$) function of $\lambda_k$ is used.

Phases can now be defined: For a phase nucleus, $x$ (an extremum in $\Acal_m$ for appropriate $m$), the distance to the nearest other nucleus was found. All points within half that distance were associated with $x$ and called its phase. This left points in ``no man's land,'' which was formally treated as another phase. For $\hat J$ they were about 40\% of the points. Nevertheless, their total $p_0$ measure was only $\sim2\!\times\!10^{-5}$.

The next step was to study the induced dynamics among the phases. An ensemble of points was started within each phase with the probability of being in a given state (within that phase) proportional to its $p_0$ measure \cite{note:p0measureinitialconditions}. I then checked the likelihood that on exiting this phase it went to a particular other phase \cite{note:nomansland}. This coarse-grained matrix of transition probabilities is designated~$\Rcg$.

\begin{figure}
\centerline{\hskip -130pt
\vbox{\includegraphics[height=.22\textheight,width=.5\textwidth]{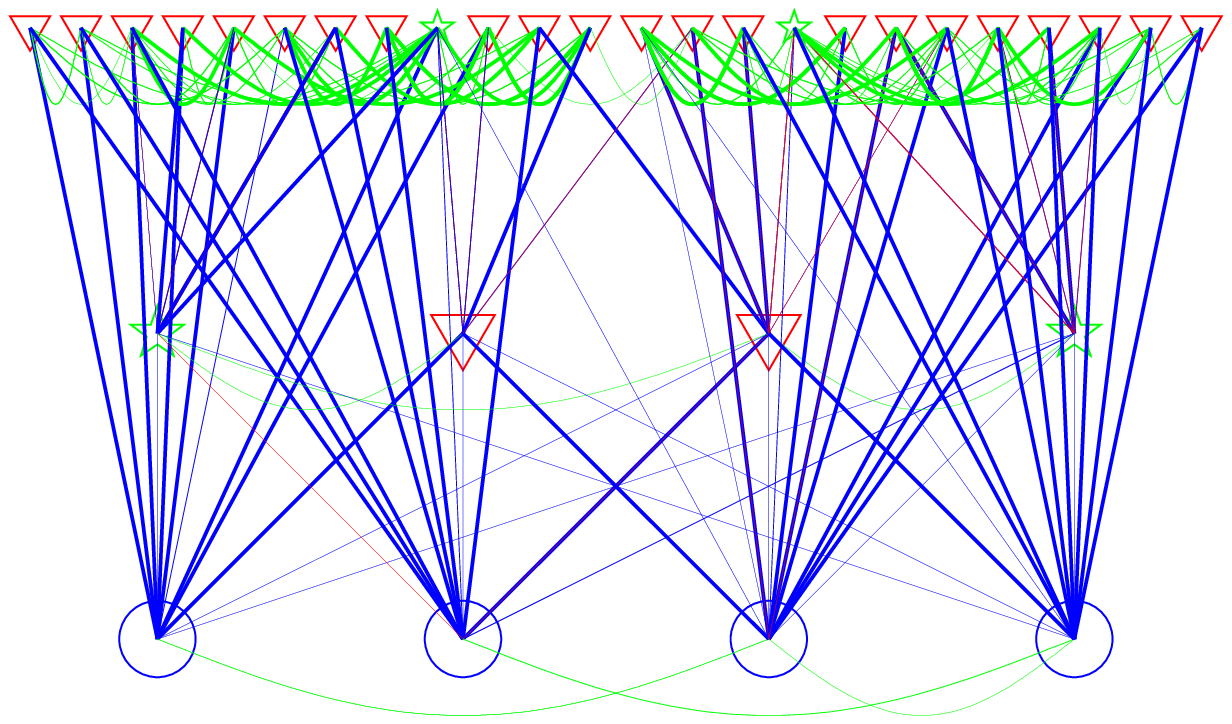}\\~
} \hskip -100pt
\includegraphics[height=.27\textheight,width=.4\textwidth]{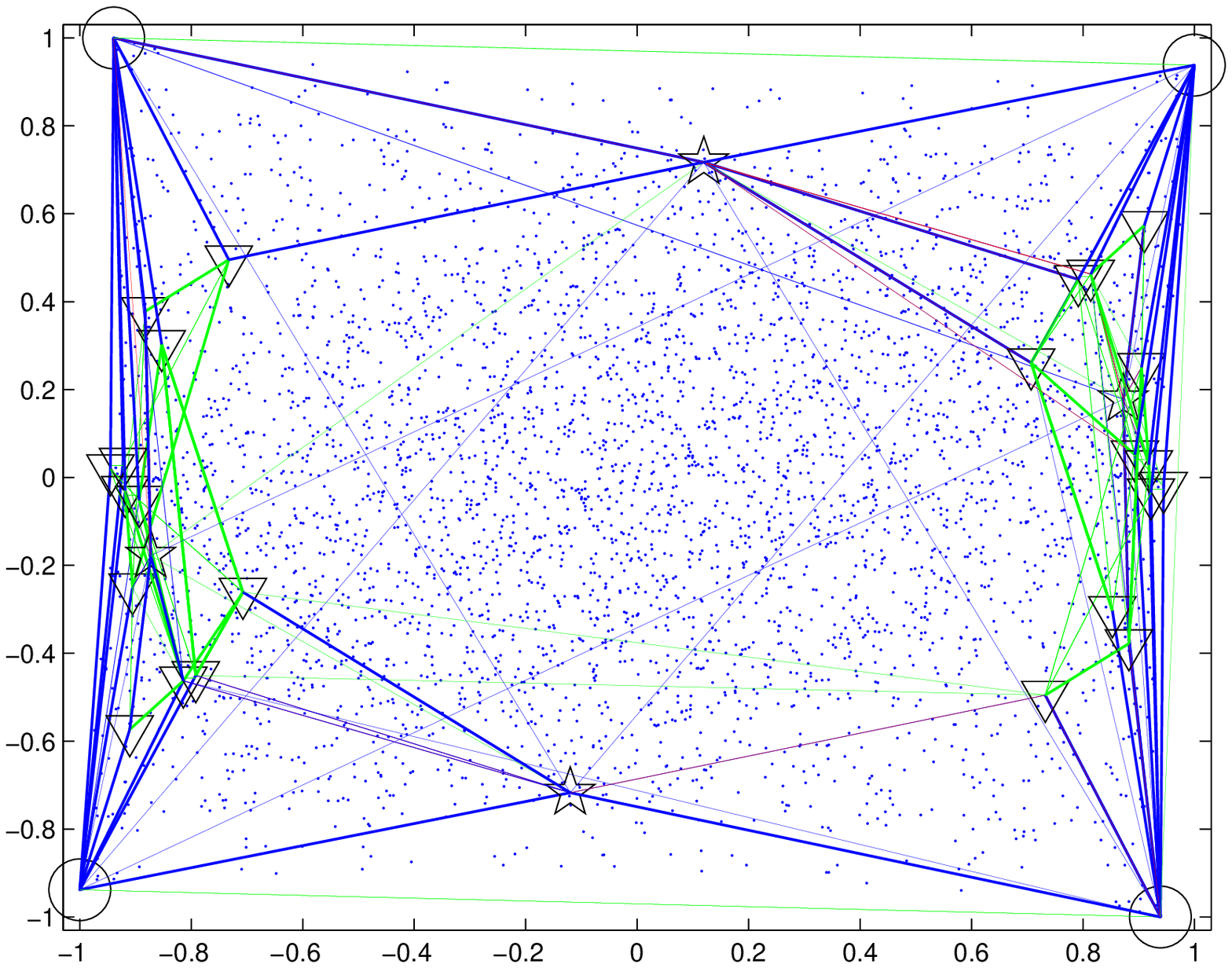}}
\caption{Representations of the matrix $\Rcg$, as explained in the text. The quench is $\hat J$. The sequence of symbols (in order of declining amount) is circle, star, base-up triangle, base-down triangle. The sequence of colors is black, red, blue, green, cyan. \label{fig:phaseconnec1}}
\end{figure}

It is in a diagrammatic representation of the flows in $\Rcg$ that one can hope to see the hierarchical structure associated with relaxation through a succession of metastable phases. Fig.\ \ref{fig:phaseconnec1} shows two such representations. In the first, the phases are sorted (based on a histogram of lifetimes) into 3 categories, long-, medium- and short-lived. On the lowest level are the longest-lived, etc. The symbol and color correspond to the amount of inflow. A phase's left-right position follows the $A_1$ values of that phase's nucleus among phases at the same lifetime level. Lines indicate transition probabilities, with line width a monotonic function of transition probability. Blue lines are flow from higher (shorter lifetime) levels to lower, red go  oppositely, and green represents lateral transitions. A perfect hierarchical structure would be almost all blue and each node would flow to only one other node. This is clearly not the case, with the principal deviations related to the col-like metastable phases.

A second display is to superimpose the flows of $\Rcg$ on the \obsrep\ for $R$ (putting a phase at the location of its nucleus). This is the second image in Fig.\ \ref{fig:phaseconnec1}. Now the symbol reflects the energy of the nucleus of the phase, and its size the lifetime. Again the breakdown in tree structure arises principally from the quasi-cols. The stable phases are in the corners, and those metastable phases that have flows to more than one side of the diagram are of intermediate lifetime and have $A_1$ near zero.

If one were to watch the time evolution of the system on this embedding, it would have the appearance of a random walk on a landscape of hills and valleys, as envisioned for example in \KK\ \cite{kobe}, who present images very much resembling ours. The advantage of using the \obsrep\ is that I do not need to resort to hamming distance (which can be a poor measure of dynamical proximity, but have a metric that is automatic and has physical significance.

\begin{figure}
\centerline{\hskip -130pt
\vbox{\includegraphics[height=.22\textheight,width=.5\textwidth]{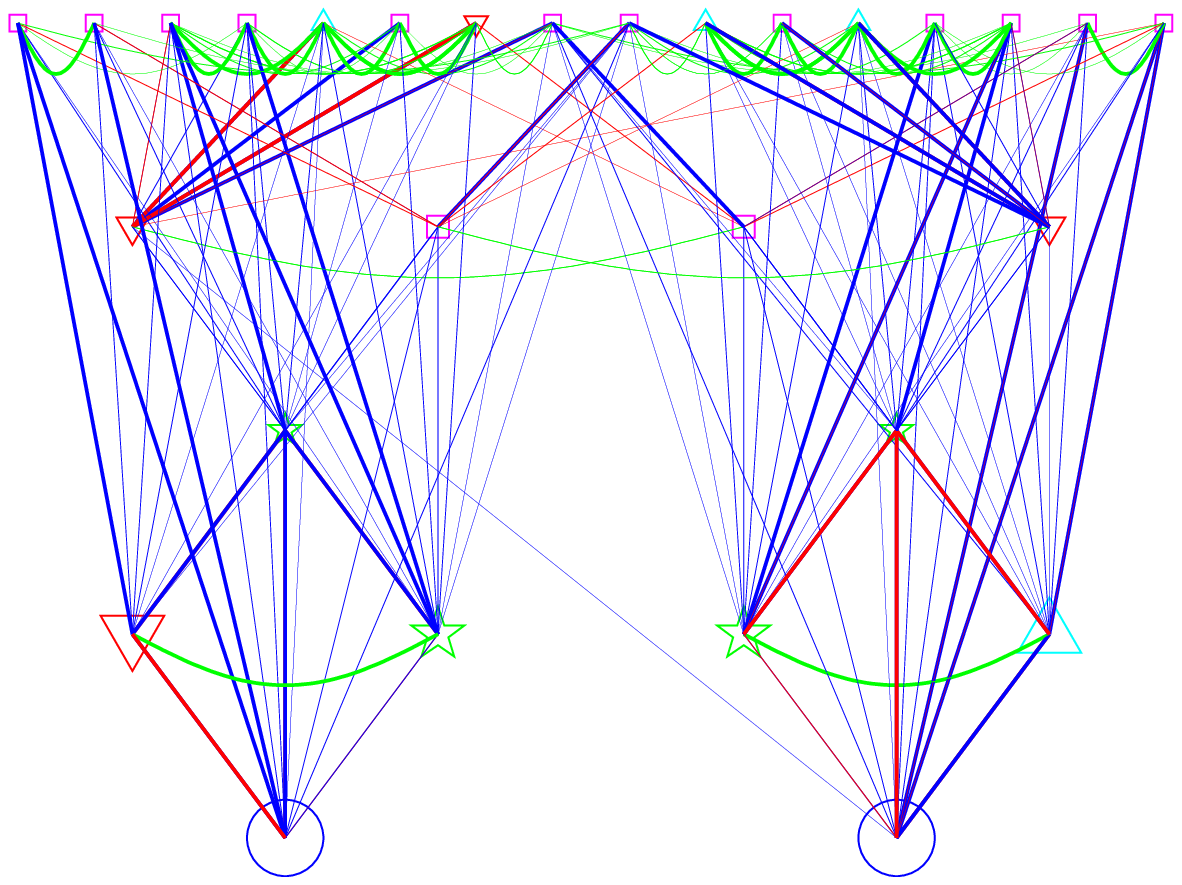}\\~
} \hskip -100pt
\includegraphics[height=.27\textheight,width=.4\textwidth]{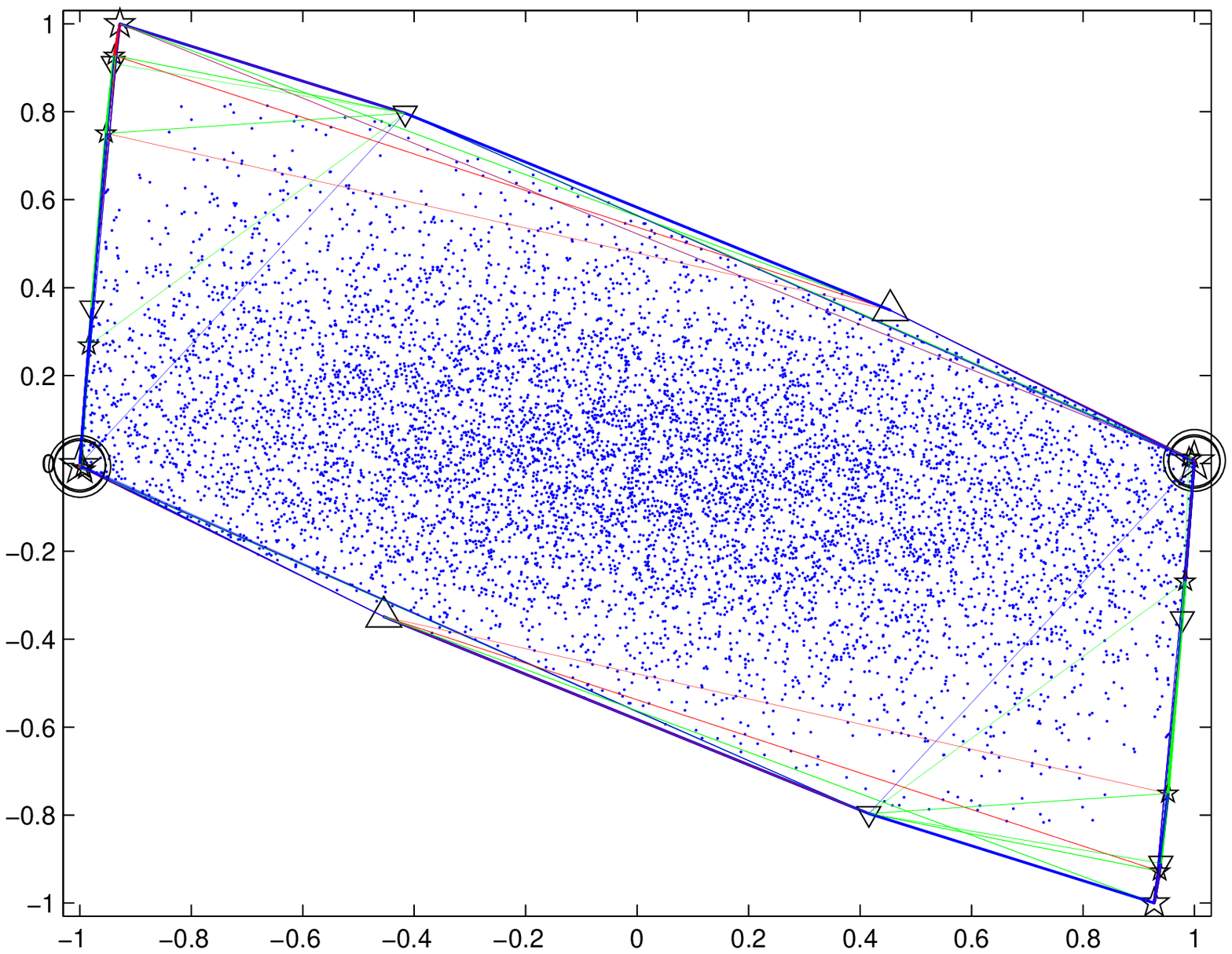}}
\caption{Representations of the matrix $\Rcg$ for a 13 spin system, as explained in the text. The same symbols, etc., are used as in the previous figure. \label{fig:phaseconnec2}}
\end{figure}

The variation of these figures with quench can be seen in Fig.\ \ref{fig:phaseconnec2}, illustrating $\Rcg$ for a 13-spin system. There are no isolated local minima (there are equal energy pairs differing by a spin flip), nor do the local-minima exhaust the extrema used. Here a more tree-like structure occurs. The cols (the most important being the base-down triangles at $A_1\approx\pm0.45$) are both of lesser weight and have larger $|A_1|$ values. This also is evident in Fig.\ \ref{fig:phaseconnec2}a, since the principal phases that feed both sides have relatively short lifetimes.

Returning to $\hat J$, I interpret the transients in Fig.\ \ref{fig:structureq}. I have already noted that the central peak persists for long times because the product of points in the stable phases is zero for several combinations. In the time-200 histogram the largest peaks away from 0 or $\pm1$ are at $q=\pm1/3$. Aside from combinations of low probability these involve \textit{exclusively} the two phases at $(A_1,A_2)\approx\pm(0.12,0.72)$, which are the stars in Fig.\ \ref{fig:phaseconnec1}b (and are extrema of $\Acal_6$). Indeed, going to our lifetime data, there are only 4 phases with lifetimes on the order of hundreds. The other two are located at $(A_1,A_2)\approx\pm(0.87,0.18)$ (referred to earlier as metastable phases). Their inner product with the stable states however is either 0 or $\pm1$. I have thus identified the phases giving rise to a specific peak in the Parisi $q$-distribution.

%%%%%%%%%%%%%%%%%%%%%%%%%%%%%%%%%%%%%%%%%%%%%%%%%%%%%%%%%%%%%%%%%%%
\prlsection{Summary and prospects}
%%%%%%%%%%%%%%%%%%%%%%%%%%%%%%%%%%%%%%%%%%%%%%%%%%%%%%%%%%%%%%%%%%%
The \obsrep\ for the mean field spin glass (and for other systems as well) provides a continuous embedding of a discrete state space. In this representation, the distance between points reflects their dynamical proximity and the extrema of the convex hull of the set of points correspond to phases, both stable and metastable. Moreover, the barycentric coordinates of a point located in the interior of the convex hull provide probabilities for the asymptotic arrival of this point in one or another of the phases.

\begin{figure}
\includegraphics[height=.3\textheight,width=.43\textwidth]{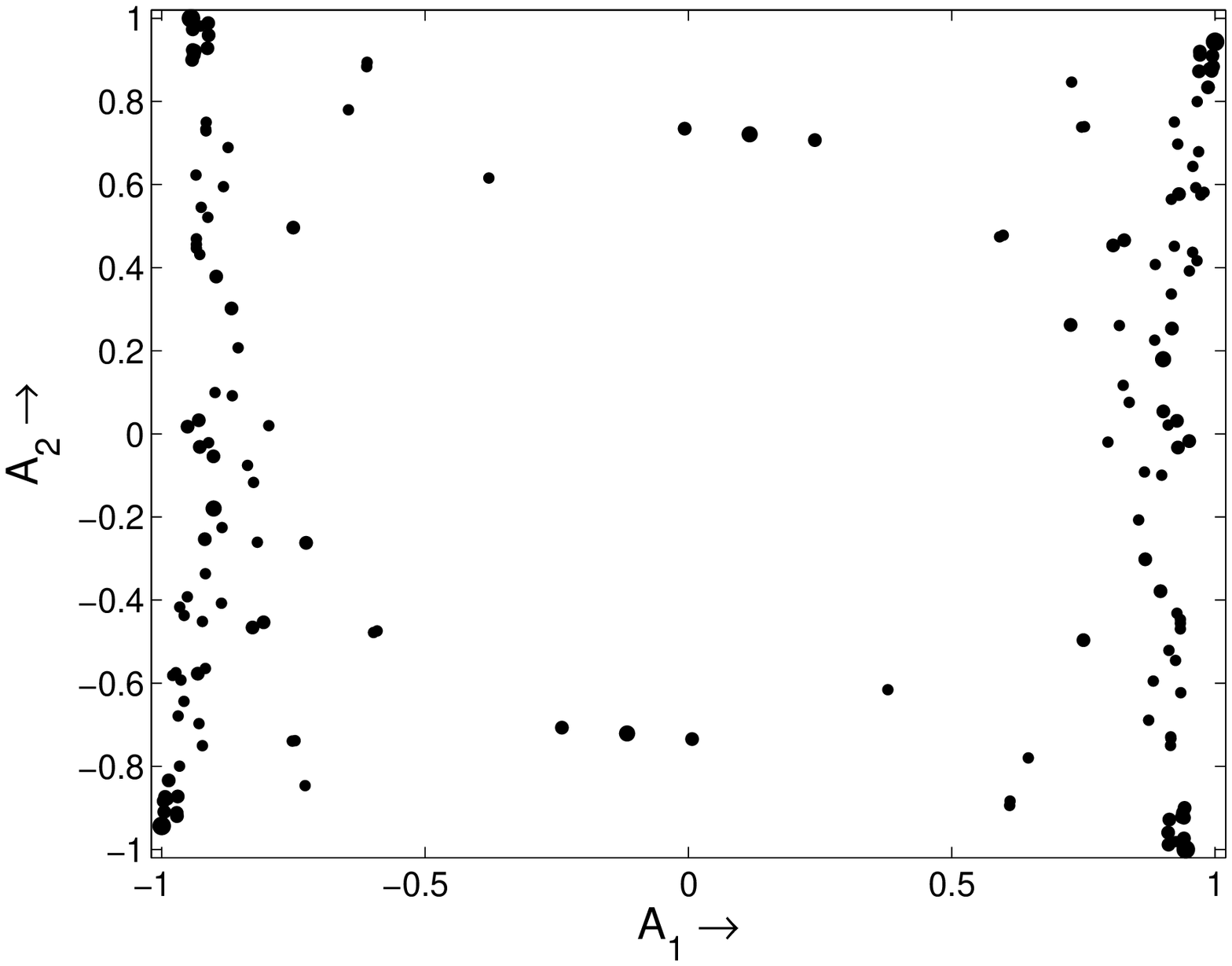}
\caption{\Obsrep\ for the same quench (and the same temperature) as Fig.\ \ref{fig:obsrepJd2}.}\label{fig:reduced}
\end{figure}

An obstacle to the implementation of this representation is the growth of the state space with $N$, the number of spins. There are two reasons, however, that it should prove possible to go well beyond the $N$'s used in this article. First, my own resources, both with respect to programming and hardware are modest. But more important is the possibility of focusing on the \textit{significant} states. This is done, for example, in Ref.\ \cite{kobe}, where for the 4\x4\x4 lattice spin glass ($2^{64}$ spin configurations) they restrict themselves to the first 3 levels, comprising $1635796 \approx2^{20.6}$ states, a reduction of more than 40 powers of 2. To test whether such a reduction would affect the \obsrep, I restricted the matrix $R$ associated with the quench $\hat J$ to the first 4 energy levels. This gave 162 states in place of 4096. The associated \obsrep\ is shown in Fig.\ \ref{fig:reduced}. Comparison with Fig.\ \ref{fig:obsrepJd2} shows that the restriction has left the important information intact. This justifies optimism that the 3\x3\x3 cube should yield to this analysis, and perhaps larger systems as well.

%%%%%%%%%%%%%%%%%%%%%%%%%%%%%%%%%%%%%%%%%%%%%%%%%%%%%%%%%%%%%%%%%%%%%%%%%%%
\begin{acknowledgments}
This work is based on a continuing collaboration with B. Gaveau, and although the present article is my doing, it has involved extensive discussions with him, for which I express my gratitude. I also thank A. Billoire, B. Duplantier and J. Kurchan for helpful discussions. I acknowledge the hospitality of the SPhT, CEA Saclay and the MPI for the Physics of Complex Systems, Dresden, where much of this work was performed. This research was supported by NSF Grant PHY 0555313.
\end{acknowledgments}
%%%%%%%%%%%%%%%%%%%%%%%%%%%%%%%%%%%%%%%%%%%%%%%%%%%%%%%%%%%%%%%%%%%%%%%%%%%
%\bibliography{SKObsRep,c:/LS/Articles/InPrep/LSS_Comp06}

\begin{thebibliography}{16}
\bibitem[{not({\natexlab{a}})}]{note:p0measureinitialconditions}
\bibinfo{note}{Even if a phase is differently populated, it rapidly goes to a
  restricted Boltzmann distribution (i.e., $\propto p_0$).}
\bibitem[{not({\natexlab{b}})}]{note:nomansland}
\bibinfo{note}{For points entering ``no man's land'' a different rule was
  employed. First entry was not counted in $\Rcg$, and the transition target
  was taken to be the phase entered on the subsequent step. If this was still
  not a phase, then ``no man's land'' was taken to be the target.}
\end{thebibliography}
%%%%%%%%%%%%%%%%%%%%%%%%%%%%%%%%%%%%%%%%%%%%%%%%%%%%%%%%%%%%%%%%%%%%%%%%%%%%%

\end{document}